%% file: main.tex
\documentclass[usenatbib]{mnras}
\usepackage{graphicx}
\usepackage{amsmath}
\usepackage{amssymb}
\usepackage{url}
\usepackage{CJKutf8}
\usepackage{array}
\usepackage{threeparttable}

\usepackage[T1]{fontenc}
\usepackage{orcidlink}

\input{new_commands}

\makeatletter
\newcommand*{\@rowstyle}{}

\newcommand*{\rowstyle}[1]{
\gdef\@rowstyle{#1}%
\@rowstyle\ignorespaces%
}

\newcolumntype{=}{
>{\gdef\@rowstyle{}}%
}

\newcolumntype{+}{
>{\@rowstyle}%
}

\newcolumntype{C}[1]{>{\centering\arraybackslash}p{#1}}
\makeatother

\pdfoutput=1

\title[Galaxy Constraints from 21-cm PS]{Fisher matrix forecasts on the astrophysics of galaxies during the epoch of reionisation from the 21-cm power spectra}

\author[Balu et al.]{Sreedhar Balu\orcidlink{0000-0002-5281-5151},$\!^{1, 2}$\thanks{E-mail:bsreedhar@student.unimelb.edu.au}
Bradley Greig\orcidlink{0000-0002-4085-2094},$\!^{1, 2}$
J. Stuart B. Wyithe\orcidlink{0000-0001-7956-9758} $\!^{1, 2}$
\\
$^{1}$School of Physics, University of Melbourne, Parkville, VIC 3010, Australia\\
$^{2}$ARC Centre of Excellence for All Sky Astrophysics in 3 Dimensions (ASTRO 3D)
}
\date{Accepted XXX. Received YYY; in original form ZZZ}

\pubyear{2023}

\begin{document}
\label{firstpage}
\pagerange{\pageref{firstpage}--\pageref{lastpage}}
\begin{CJK}{UTF8}{gkai} 
\maketitle
\end{CJK}

\begin{abstract}
\noindent
The hyperfine 21-cm transition of neutral hydrogen from the early Universe ($z>5$) is a sensitive probe of the formation and evolution of the first luminous sources. Using the Fisher matrix formalism we explore the complex and degenerate high-dimensional parameter space associated with the high-$z$ sources of this era and forecast quantitative constraints from a future 21-cm power spectrum  (21-cm PS) detection. This is achieved using $\textsc{Meraxes}$, a coupled semi-analytic galaxy formation model and reionisation simulation, applied to an $N$-body halo merger tree with a statistically complete population of all atomically cooled galaxies out to $z\sim20$. Our mock observation assumes a 21-cm detection spanning $z \in [5, 24]$ from a 1000 h mock observation with the forthcoming Square Kilometre Array and is calibrated with respect to ultraviolet luminosity functions (UV LFs) at $z\in[5, 10]$, the optical depth of CMB photons to Thompson scattering from \textit{Planck}, and various constraints on the IGM neutral fraction at $z > 5$. In this work, we focus on the X-ray luminosity, ionising UV photon escape fraction, star formation and supernova feedback of the first galaxies. We demonstrate that it is possible to recover 5 of the 8 parameters describing these properties with better than $50$ per cent precision using just the 21-cm PS. By combining with UV LFs, we are able to improve our forecast, with 5 of the 8 parameters constrained to better than $10$ per cent (and all below 50 per cent).
\end{abstract}


\section{Introduction}

Following recombination, the early Universe entered the cosmic Dark Ages characterised by a neutral intergalactic medium (IGM) and the absence of luminous sources. The formation of the first stars and galaxies ushered in the era of the Cosmic Dawn. The intense ionising ultraviolet (UV) radiation, characteristic of the young and massive stars, as well as X-rays (possibly from high-mass X-ray binaries, e.g. \citealt{MesingerXRay}) impacted the thermal and ionisation state of the IGM. This period was brought to its conclusion in the Epoch of Reionisation (EoR) when the UV photons ionised the neutral hydrogen (\hi) rendering it transparent to UV photons.  

Considerable effort has been expended in the last few decades to unravel the complex physics of this period (see \cite{EoRbook} and references therein). The forbidden 21-cm hyperfine transition signal is well-suited for this purpose because of its extreme sensitivity to the formation and evolution of the first stars and galaxies ($z\lesssim 30$) as well as the various feedback mechanisms in the early Universe. Though the ultimate goal of 21-cm interferometric experiments is to map the tomography of the IGM as a function of frequency (redshift) the first set of observations will be statistical in nature. The 21-cm signal is observed as a brightness temperature against a background source (which is almost always assumed to be the cosmic microwave background (CMB) radiation; e.g. \citealt{BibleReview, StuartReview, PritchardReview}). The 21-cm power spectrum (21-cm PS) quantifies the fluctuations in the brightness temperature of the 21-cm signal across the sky. Though a detection has yet to be made, current experiments such as the Murchison Widefield Array \citep[MWA\footnote{\href{https://www.mwatelescope.org/}{www.mwatelescope.org}};][]{MWA_main}, LOw Frequency ARray \citep[LOFAR\footnote{\href{https://www.astron.nl/telescopes/lofar}{www.astron.nl/telescopes/lofar}};][]{LOFAR_main}, Hydrogen Epoch of Reionization Array \citep[HERA\footnote{\href{http://reionization.org/}{reionization.org/}};][]{HERA_main} have already begun setting upper limits on the 21-cm PS \citep[see][]{LOFAR_limits, MWA_limits, HERA_limits}. The upcoming Square Kilometre Array \citep[SKA\footnote{\href{https://www.skao.int/}{www.skao.int}};][]{SKA_main} will revolutionise 21-cm EoR cosmology with its unprecedented sensitivity.

The amplitude and shape of the 21-cm PS are extremely sensitive to the thermal and ionisation state of the IGM \citep{BarkanaReview} and hence the astrophysics of early galaxy formation and evolution \citep{Dragons5, EoRbook}. It is therefore imperative that realistic and efficient models \citep[see][for a recent review]{GnedinMadau} are available to interpret current and upcoming observations. Despite significant progress in this direction, there is considerable uncertainty about the properties of the underlying source populations of these models.

In this paper, we ask: \textit{What can we learn about the underlying physical processes driving reionisation from a successful detection of the 21-cm PS?} Simulating reionisation requires large volumes ($\gtrsim200 \oneh$ Mpc according to \cite{Iliev2014, Kaur2020} for convergent reionisation and 21-cm statistics). 
As a result, a number of models have been developed to make large-scale but computationally efficient realisations of the early Universe \citep[e.g.][]{Battaglia2013, GRIZZLY, Simfast, SCRIPT, 21cmFASTv3}. To investigate the utility of the 21-cm PS as a probe of galaxy formation, MCMC methods have been utilised to place constraints on the parameterised properties (e.g. UV and X-ray) of population II \& III star-forming galaxies 
\citep[e.g.][]{CMMC1, CMMC2, Park2019, Yuxiang_tale1, Yuxiang_tale2, Maity_2022, Bevins_2023}. 

In this work, we use \meraxes \citep{Dragons3} - a semi-analytic model (SAM) of galaxy formation \citep[see][for a recent review]{SomervilleReview} and evolution self-consistently coupled to a reionisation model\footnote{For alternate approaches see \citep[e.g.][]{RSAGE, Visbal2020, Hutter2021, POLAR}} - to forecast constraints on astrophysical properties of the early galaxies. 

Unlike semi-numerical models (for example, \citealt{21cmFAST}, \citealt{SCRIPT}), which focus on population-averaged quantities (i.e. these models generally have no galaxies),  \meraxes provides a  realistic population of galaxies as sources of photons. \meraxes incorporates a detailed, physically motivated galaxy formation and evolution model that includes baryonic infall, gas cooling, star formation, supernova feedback, active galactic nuclei feedback, galaxy mergers, etc. Another essential feature of \meraxes, important for its application to reionisation, is the simultaneous processing of all the galaxies in the simulation volume, thereby enabling the spatial coupling of reionisation to galaxy evolution. 

By efficiently coupling the reionisation of the IGM via a modified version of \cmfast \citep{21cmFAST, 21cmFASTv3} and an underlying galaxy population sourced from a dark matter only $N$-body simulation, \meraxes is thus well-suited to explore the underlying parameter space of the complex astrophysics of this era. We deploy \meraxes on a $210\oneh$ Mpc cosmological simulation which resolves all the atomically cooled galaxies down from $z\sim20$. The large volume and high mass resolution of our simulation make an MCMC analysis using \meraxes prohibitively expensive computationally\footnote{Recently, \cite{Dragons21} utilised  \meraxes to place constraints on the UV escape fraction of the early galaxies using existing high-$z$ observations within an MCMC framework. This was only feasible owing to (i) not considering the thermal state of the IGM and (ii) using a simulation volume 30 times smaller with a 6 times lower mass resolution for the galaxies.}.

Using a Fisher matrix analysis, we forecast the constraints on a total of 8 astrophysical parameters in our model that directly control the X-ray luminosity, UV escape fraction, star formation rates and supernova feedback of the galaxies of the high-$z$ Universe. Focusing on the upcoming SKA1-low, we forecast constraints from the 21-cm PS before exploring the improvements available when combining information from the UV LFs. 

The paper is organised as follows: In section \ref{sec:2} we introduce our \textit{N}-body simulation (section \ref{sec:genpsim}), galaxy SAM (section \ref{sec:meraxes}) and the reionisation model (section \ref{sec:reion}). We also introduce our set of eight astrophysical parameters in this section. In section \ref{sec:sense} we introduce our mock observation, in section 4 we describe the Fisher Matrix formalism and we analyse our results in section 5. Our simulations use the best-fit parameters from the \cite{Planck2015}: $h=0.6751$, $\Omega_{\rm m}=0.3121$, $\Omega_{\rm b}=0.0490$, $\Omega_\Lambda=0.6879$, $\sigma_8=0.8150$, and $n_s=0.9653$. All quantities quoted are in comoving units unless otherwise stated.

\section{Simulating the 21-cm signal}\label{sec:2}

We give a brief review of our underlying dark matter-only N-body simulation and the \meraxes SAM in this section.  We focus on a subset of the free parameters in our model that directly impact the star formation, supernova feedback, UV escape fraction, and X-ray luminosity of the galaxies.

\subsection{N-body merger trees}\label{sec:genpsim}

We use the L210\_N4320 dark matter-only \textit{N}-body simulation of the \textit{Genesis} suite of simulations (Power et al. in prep). Containing $4320^3$ particles in a $210^3~h^{-3}$ Mpc$^3$ volume, the simulation has a mass resolution of $\sim5\times10^8~\oneh~\Msun$. L210\_N4320 was run with the \textsc{SWIFT} \citep{SWIFT} cosmological code, using the  \textsc{VELOCIraptor} \citep{VELOCIRAPTOR} halo identifier and merger trees generated using \textsc{TreeFrog} \citep{TREEFROG}. The simulation consists of 120 snapshots between redshifts 30 and 5. 

In order to resolve haloes down to the atomic cooling limit at $z=20$, the L210\_N4320 simulation was augmented using the Monte-Carlo algorithm-based code $\dforest$ \citep{DARKFOREST} achieving an effective halo mass resolution of $\sim 2 \times 10^7 ~\oneh~\Msun$ (\genpsim of \citealt{Balu2022}). \dforest achieves this by sampling from a conditional mass function based on the extended Press-Schechter theory \citep{EPS1, EPS2, EPS3} that has been modified to match the halo mass functions from \textit{N}-body simulations. These new low-mass haloes are introduced into the simulation volume and appended to the merger trees. Importantly, \dforest also assigns positions to these newly added haloes based on the local halo density field \citep{Ahn2015} and the linear halo bias \citep{Tinker2010}. For the rest of this work, we use $\genpsim$ as our fiducial simulation.

\subsection{A realistic galaxy population from \texorpdfstring{$\textsc{Meraxes}$}{Lg}}

\label{sec:meraxes}
\begin{table*}
    \centering
    \begin{tabular}{c c c c c c c}
    \hline
    \vspace{0.1cm}\\
    Parameter & Description & Equation & Fiducial  & 21-cm PS alone & UV LF alone & 21-cm PS \& UV LF\\ 
     &  &  & value & 1-$\sigma$ values & 1-$\sigma$ values & 1-$\sigma$ values  \\
    &  &  &  &  (\% uncertainty) &  (\% uncertainty) \\
    \vspace{0.1cm}\\
    \hline
    \vspace{0.1cm}\\
    $\rm{log}_{10}\left(\dfrac{\textit{L}_{\rm X}}{\rm SFR}\right)$ & X-ray luminosity per SFR & \ref{eq:E_0} & 40.50  & 0.0043 & &  0.0027\\ 
    &  & & & ($1.1\times10^{-2}$) & -- & ($7.0\times 10^{-3}$)\\ 
    \vspace{0.15cm}\\
    $E_{0}$ & Minimum X-ray photon energy & \ref{eq:E_0} & 500.00 & 54.5058 & & 27.708 \\ 
    & & & & (10.9) & -- & (5.54)\\     
    \vspace{0.15cm}\\
    $f_{\rm esc, 0}$ & Escape fraction normalisation & \ref{eq:f_esc}  & 0.14 & 0.0092 & 0.1172 & 0.0069 \\
    & & & & (6.55) & (83.71)& (4.91)\\
    \vspace{0.15cm}\\
    $\alpha_{\rm esc}$ & Escape fraction redshift scaling & \ref{eq:f_esc}  & 0.20 & 0.1339  & 0.2552 & 0.0965 \\ 
    & & & & (66.93) & (127.6) & (48.25)\\ 
    \vspace{0.15cm}\\
    $\rm{log}_{10}(\Sigma_{\rm SF})$ & Critical mass normalisation & \ref{eq:m_crit} & -1.86 & 4.3496 & 0.9511 & 0.7401 \\
    & & & & (233.9) & (51.13) & (39.8)\\
    \vspace{0.15cm}\\
    $\rm{log}_{10}(\alpha_{\rm SF})$ & Star formation efficiency & \ref{eq:SF_law} & -1.00 & 0.0883 & 0.0661 & 0.0436\\
    & & & & (8.82) & (6.61) & (4.35)\\
    \vspace{0.15cm}\\
    $\rm{log}_{10}(\epsilon_{0})$ & Supernova ejection efficiency & \ref{eq:epsilon} & 0.19 & 0.0470 & 0.0668 & 0.0309  \\ 
    & & & & (25.1) & (35.16) & (16.56) \\ 
    \vspace{0.1cm}\\
    $\rm{log}_{10}(\eta_{0})$ & Supernova reheat efficiency & \ref{eq:eta}  & 0.84  & 0.9876  & 0.0991 & 0.0658 \\ 
    & & & & (117.03) & (11.8) & (7.803)\\ 
    \vspace{0.15cm}\\
    \hline  
    \end{tabular}
    \caption{The first column lists the free astrophysical model parameters for which we forecast the uncertainties with a Fisher matrix formalism. The next three columns give a short description, the corresponding equation in the text, and their adopted fiducial values in this work respectively. The fifth and sixth columns list the forecasted 1-$\sigma$ constraints using just the 21-cm PS and just the UV LFs respectively, and the final column gives the same for a joint analysis of both the 21-cm and the UV LFs. The fractional uncertainties are given in brackets as a percentage. We do not vary the X-ray parameters, \lxraygal and $E_0$, for the UV LFs analysis as they do not have an impact on the UV LFs.}
    \label{table:params}
\end{table*}

\meraxes \citep{Dragons3} contains detailed and physically motivated prescriptions for galaxy formation and evolution. These include, for example, gas infall into dark matter haloes from the IGM followed by its radiative cooling and subsequent star formation, as well as eventual supernova feedback and metal enrichment of the ISM. Active galactic Nuclei (AGN) feedback from central black holes of the galaxies was implemented in \cite{Dragons10} and calculations of the galaxies' UV luminosity in \cite{Dragons19}. In addition, \meraxes also has a coupled treatment of reionisation based on the \cmfast semi-numerical code \citep[][see section \ref{sec:reion} for more details]{21cmFAST, 21cmFASTv3}. In the following sections, we give detailed but non-exhaustive descriptions of the various physics implementations pertinent to our astrophysical model. Table \ref{table:params} lists the free parameters of our model along with their adopted fiducial values.

\subsubsection{Star formation prescription}\label{sec:SF}
At every snapshot, the baryonic content of a dark matter halo increases by $(1-f_{\rm mod})f_{\rm b}M_{\rm vir}$, where $f_{\rm b} = \Omega_{\rm b}/\Omega_{\rm m}$ is the baryonic fraction and $f_{\rm mod}$ (equation \ref{eq:f_mod}) is the baryon fraction modifier - set by the local IGM ionisation state from the previous snapshot -  coupling galaxy growth to the ionisation state of the IGM (see section \ref{sec:f_mod} for details), and $M_{\rm vir}$ is the halo virial mass.  This newly accreted baryonic gas is deposited to a `hot-gas reservoir' of the halo from where it cools radiatively to a `cold-gas component' from which it can form stars.

Star formation in \meraxes follows the disc stability argument of \cite{Kauffmann1996}, wherein gas participates in star formation when the cold-gas mass is higher than a critical mass ($m_{\rm crit}$) given by
\begin{equation}\label{eq:m_crit}
    m_{\rm crit} = \Sigma_{\rm SF} \bigg(\dfrac{V_{\rm max}}{100 \rm{ kms}^{-1}}\bigg)\bigg(\dfrac{r_{\rm disc}}{10 \rm{ kpc}}\bigg)\times 10^{10} \Msun,
\end{equation}
where $\Sigma_{\rm SF}$ is the critical mass normalisation, $V_{\rm max}$ is the maximum halo circular speed, and $r_{\rm disc}$ is the scale radius of the galactic disc. The new stellar mass, $\Delta M_{\rm star}$, formed in the time-step $\Delta t$ is given by
\begin{equation}\label{eq:SF_law}
    \Delta M_{\rm star} = \alpha_{\rm SF} \frac{m_{\rm cold} - m_{\rm crit}}{t_{\rm dyn, disc}}\Delta t,
\end{equation}
where $\alpha_{\rm SF}$ is the star formation efficiency, $m_{\rm cold}$ is the mass locked up in the cold-gas component, and $t_{\rm dyn, disc}$ is the dynamical time of the disc given by $r_{\rm disc}/V_{\rm max}$.

In this work, both $\Sigma_{\rm SF}$ and $\alpha_{\rm SF}$ are free parameters in our astrophysical model whose fractional uncertainty we predict in section \ref{sec:results}.

\subsubsection{Supernova feedback prescription} 
\label{sec:SNe}
The star formation rates and the stellar mass of galaxies are regulated by a number of feedback mechanisms including supernovae (SNe) feedback from evolved stars, AGN feedback, and the ionisation state of the IGM regulated by the progress of reionisation. \cite{Dragons3} showed that SNe feedback dominates over self-regulation by the EoR. In this subsection, we detail the SNe implementation in our model.

The primary impact of SNe feedback is to heat the gas reservoirs of the halo resulting in the transfer of the gas from the cold to the hot gas reservoirs and in extreme cases the removal of the gas from the hot halo (i.e. altogether from the galaxy). Our implementation, modified from \cite{Guo2011} to take advantage of our high cadence merger trees \citep[see][]{Dragons19}, is based on energy conservation. The total stellar mass going SNe $(\Delta m_{\rm new})$ at a particular snapshot of the simulation depends on both the current and previous star formation\footnote{As the largest time-step of our simulation is $\sim 16$ Myr, newly formed massive stars can go SNe in the same time-step and less massive stars can survive for a few snapshots.}. To account for this, we track the stars formed in the present and four previous snapshots. The stellar mass going SNe at a particular snapshot is calculated as a weighted average star formation history 
\begin{equation}\label{eq:m_new}
\Delta m_{\text {new }}=\dfrac{\int_t^{t+\Delta t} \mathrm{ d} t^{\prime} \int_0^{\infty} \mathrm{d} \tau \frac{\mathrm{d} \varepsilon}{\mathrm{d} \tau} \psi\left(t^{\prime}-\tau\right)}{\int_0^{\infty} \mathrm{d} \tau \frac{d \varepsilon}{\mathrm{d} \tau}},
\end{equation}
where $\Delta t$ is the simulation time-step, $d\varepsilon/d\tau$ is the rate of energy release per unit stellar mass via Type-II SN by stars within age $\tau$ and $\tau + d\tau$, and  $\psi(t)$  is the star formation rate at time $t$ of the galaxy.  We calculate $d\varepsilon/d\tau$ from \textsc{STARBURST99} \citep{Leitherer1999, Leitherer2010, Leitherer2014} accounting for both metallicity as well as a \cite{Kroupa2002} initial mass function. The energy released by SNe can then be calculated as 
\begin{equation}
\Delta E_{\mathrm{SN}}=\epsilon \int_t^{t+\Delta t} \mathrm{ d} t^{\prime} \int_0^{\infty} \mathrm{d} \tau \frac{\mathrm{d} \varepsilon}{\mathrm{d} \tau} \psi\left(t^{\prime}-\tau\right),
\end{equation}
where $\epsilon$ is the energy coupling efficiency.

The amount of gas reheated from the cold gas reservoir ($\Delta m_{\rm reheat}$) or ejected altogether from the halo ($\Delta m_{\rm eject}$) is calculated as
\begin{equation}\label{eq:m_reheat}
\Delta m_{\text {reheat }}=\left\{\begin{array}{ll}
\eta \Delta m_{\text {new }}, & \Delta E_{\mathrm{SN}} \geq \Delta E_{\mathrm{hot}} \\\\
\dfrac{\Delta E_{\mathrm{SN}}}{1 / 2 V_{\mathrm{vir}}^2}, & \Delta E_{\mathrm{SN}}<\Delta E_{\mathrm{hot}}
\end{array}\right.,
\end{equation}
and
\begin{equation}
\Delta m_{\mathrm{eject}}=\frac{\Delta E_{\mathrm{SN}}-\Delta E_{\mathrm{hot}}}{1 / 2 V_{\mathrm{vir}}^2},
\end{equation}
where 
\begin{equation}\label{eq:E_hot}
\Delta E_{\mathrm{hot}}=\frac{1}{2} \eta \Delta m_{\mathrm{new}} V_{\mathrm{vir}}^2,
\end{equation}
$\eta$ is the mass loading factor and $V_{\rm vir}$ is the halo virial velocity. Following \cite{Dragons19}, $\epsilon$ and $\eta$ are  implemented as 
\begin{equation} \label{eq:epsilon}
    \epsilon = \begin{cases}
    \epsilon_0 \bigg(\dfrac{1 + z}{4}\bigg) \bigg( \dfrac{V_{\rm max}}{70 {\rm  km s^{-1}}}\bigg)^{-1}, & V_{\rm max} \geq 70 {\rm km s^{-1}} \\ \\
    \epsilon_0 \bigg(\dfrac{1 + z}{4}\bigg) \bigg(\dfrac{V_{\rm max}}{70 {\rm km s^{-1}}}\bigg)^{-3.2}, & V_{\rm max} < 70 {\rm km s^{-1}}\\
  \end{cases},
\end{equation}
and,
\begin{equation} \label{eq:eta}
    \eta = \begin{cases}
    \eta_0 \bigg(\dfrac{1 + z}{4}\bigg)^{2}\bigg(\dfrac{V_{\rm max}}{60\rm{ km s^{-1}}}\bigg)^{-1}, & V_{\rm max} \geq 60 {\rm  km s^{-1}} \\ \\
    
    \eta_0 \bigg(\dfrac{1 + z}{4}\bigg)^{2}\bigg(\dfrac{V_{\rm max}}{60 {\rm km s^{-1}}}\bigg)^{-3.2}, & V_{\rm max} < 60 {\rm km s^{-1}} \\
  \end{cases}
\end{equation}
respectively, where $V_{\rm max}$ is the maximum halo circular velocity, $\epsilon_0$ is the supernova energy coupling normalisation, and $\eta_0$ is the supernova ejection efficiency. 

We forecast the fractional uncertainty on both $\epsilon_0$ and $\eta_0$ and summarise their values in Table \ref{table:params}.

\subsubsection{Escape fraction of the UV ionising photons}
The fraction of ionising photons escaping into the IGM from galaxies plays an important role in regulating the ionisation fraction and morphology. The model parameters that directly impact the ionisation state have a pronounced effect on the 21-cm PS. Following \cite{Dragons3} we employ an escape fraction $(f_{\rm esc})$ prescription for galaxies that is solely redshift dependent  \citep[though see e.g.][]{Kimm2017, Yeh2023}. This results in an $f_{\rm esc}$ that is skewed towards higher $z$, motivated by two factors. First, it is easier for the photons to climb out of the shallower potential well of low-mass high-z galaxies. Second, early galaxies are also characterised by less dust attenuation compared to their low-redshift counterparts.
r\begin{equation}\label{eq:f_esc}
    f_{\rm esc} = f_{\rm esc, 0}\bigg(\frac{1+z}{6}\bigg)^{\alpha_{\rm esc}}
\end{equation}
We allow both the escape fraction normalisation \fnorm and redshift scaling $\alpha_{\rm esc}$ to be free parameters in this work.

\subsection{Evolution of the IGM}\label{sec:reion}
The thermal evolution and ionisation state of the IGM follows \cmfast \citep{21cmFAST, 21cmFASTv3}, modified to take advantage of our realistic galaxy population. \meraxes sources the dark matter density and velocity fields from the underlying \textit{N}-body simulation. We begin by gridding our simulation volume and assigning the galaxies to voxels based on their positions. In the present work, motivated by the typical $\hii$ bubble sizes during the EoR (\citealt{Furlanetto2004}, \citealt{Stu2004}), we divide our simulation into $1024^3$ voxels corresponding to a cell dimension of $\sim 0.21~\oneh$ Mpc. We give a brief summary of our method in the following subsections.

\subsubsection{Thermal state of the IGM}
\label{sec:Ts}
The thermal state of a \hi cloud, characterised by the spin temperature $T_{\rm S}$, depends upon the radiation that is impinging upon it. Even though $T_{\rm S}$ is influenced by both UV and X-ray photons, the latter has a considerably more pronounced impact \citep[e.g.][]{MesingerXRay} and is computed as 
\begin{equation}
    T_{S}^{-1} = \dfrac{T_{\rm CMB}^{-1} + x_{\alpha} T_\alpha^{-1} + x_{\rm c} T_{\rm{K}}^{-1}}{1 + x_{\alpha} + x_{\rm{c}}},
\end{equation}
where $T_{\rm CMB}$ is the temperature of the CMB, $T_{\rm K}$ is the gas kinetic temperature,  $T_\alpha$, which we take to be equal to $T_{\rm K}$, is the colour temperature, $x_\alpha$ is the Wouthuysen-Field (WF) coupling constant (\citealt{Wouthuysen1952} \& \citealt{Field1958}), and $x_{\rm c}$ is the collisional coupling coefficient which accounts for the  collisional coupling of the gas with other \hi atoms, free electrons, and free protons. Here, we summarise only the implementation of the X-rays (which largely impacts the $T_{\rm K}$) and refer the reader to \cite{Balu2022} for further details.

An X-ray photon with energy $E_e$ when emitted at redshift $z^{'}$, redshifts to energy $E = E_e (1 + z) / (1 + z^{'})$ at redshift $z$. We compute the comoving X-ray emissivity $\epsilon_X(\boldsymbol{x}, E_e, z')$ in the emitted frame at location $\boldsymbol{x}$ as 
\begin{equation}
\epsilon_X(\boldsymbol{x}, E_e, z') = (L^{'}_{\rm X}/\rm{SFR})\times \rm{SFRD}(\boldsymbol{x}, z'),
\end{equation}
where $L^{'}_{\rm X}/{\rm SFR}$  is the specific X-ray luminosity per SFR\footnote{Note that this SFR is an average quantity of all the galaxies in a grid and is different from the $\psi$ in equation (3), which is for each galaxy.} and SFRD$(\boldsymbol{x}, z')$ is the star formation rate density. The $SFRD$ grid depends on the local galaxy population, enabling us to couple galaxy evolution with the thermal state of the IGM. $L^{'}_{\rm X}/{\rm SFR}$ is implemented as a power-law of the X-ray energy as 
\begin{equation}\label{eq:alpha_X}
        L^{'}_{\rm X}/\rm{SFR} \propto \textit{E}^{-\alpha_\textit{X}},
\end{equation}
where $\alpha_{X}$ is the X-ray spectral index and is normalised as 
\begin{equation}
\label{eq:E_0}
L_{\rm X}/\rm{SFR} = \int_{\textit{E}_0}^{2 keV} d\textit{E}_e \textit{L}^{'}_{X}/{\rm SFR}.
\end{equation}
\lxraygal is one of the free parameters of our model and has units [erg $\rm s^{-1} \Msun^{-1} \rm{yr}$]\footnote{Note, our \lxraygal is equivalent to  $L_{\rm X<2 keV}/\rm{SFR}$ in \cmfast.}. We impose an upper limit of $2$ keV since X-ray photons with higher energies have mean-free paths longer than the Hubble length and are unlikely to impact $T_{\rm S}$ \citep{McQuinn2012, Das2017}. The lower limit $E_0$ is motivated by the absorption of low-energy X-rays within the galaxy itself.

We forecast constraints on both $E_0$ and \lxraygal in this work and their fiducial values are summarised in Table \ref{table:params}.

\subsubsection{Ionisation state of the IGM}\label{sec:f_mod}
\label{sec:xhi}
We compute the ionisation ($x_{\hii}$) grid by employing an excursion-set formalism \citep{Furlanetto2004} by comparing the total number of ionising photons to the combined number of neutral atoms and recombinations within spheres of decreasing radii. Grid voxels inside a sphere of radius $R$ with centre $\boldsymbol{x}$ and at redshift $z$ are deemed to be ionised if
\begin{equation}\label{eq:excursion}
    N_{\rm b*} (\boldsymbol{x}, z | R) N_\gamma f_{\rm esc} \geq  N_{\rm atom} (\boldsymbol{x}, z | R) (1 + \bar{n}_{\rm rec}) (1 - \bar{x}_e),
\end{equation}
where $N_{\rm b*} (\boldsymbol{x}, z | R)$ is the cumulative number of stellar baryons, $N_\gamma$ is the average number of ionising photons per stellar baryon, and $f_{\rm esc}$ is their escape fraction. The left-hand side of equation \ref{eq:excursion} thus gives the total number of ionising photons in the volume. On the right-hand side, $N_{\rm atom} (\boldsymbol{x}, z | R)$ is the total number of baryons, $(1 + \bar{n}_{\rm rec})$ accounts for the recombinations inside Lyman limit systems \citep[see][]{Sobacchi2014}, and $(1 - \bar{x}_e)$ accounts for secondary ionisations caused by the X-ray photons, giving the number of neutral $\hi$ atoms in the same volume. We fix the maximum value of $R$ to be 50 Mpc \citep[see][]{Songaila2010,Becker2021}.

The number of stellar baryons, $N_{\rm b*}$, is inferred from the stellar mass that is dependent on the star formation histories of all star-forming galaxies. We compute a $f_{\rm mod}$ for each galaxy based on their local $x_{\hi}$ value, enabling us to couple a galaxy's growth to its local IGM ionisation state. In the next snapshot, the amount of fresh gas accreting onto the galaxy is modulated by $f_{\rm mod}$ which is calculated, following \cite{Sobacchi2013}, as
\begin{equation}\label{eq:f_mod}
    f_{\rm mod} = 2^{-M_{\rm filt}/M_{\rm vir}},
\end{equation}
where $M_{\rm filt}$ is the `filtering mass', corresponding to when the baryon fraction of the host halo is half of the cosmic IGM value and is given by

\begin{equation}
M_{\mathrm{filt}}=M_0 J_{21}^a\left(\frac{1+z}{10}\right)^b\left[1-\left(\frac{1+z}{1+z_{\mathrm{ion}}}\right)^c\right]^d,
\end{equation}

where $z$ is the current redshift, $z_{\rm ion}$ is the redshift when the voxel was first ionised, and $J_{21}$ is the local average UVB. The values $[M_0, a, b, c, d]$ are taken from \citep{Sobacchi2013}  (see \citealt{Dragons3} for more details). In this manner, by coupling galaxy growth to the local ionisation state as well as the local UV ionising background, \meraxes enables a self-consistent reionisation scenario. See \cite{Dragons5} for an exploration of the self-consistent and UV background regulated reionisation within \meraxes. 

\subsection{21-cm brightness temperature}
A radio telescope measures the brightness temperature ($\delta T_b$) of a \hi cloud, which is the offset of the spin temperature ($T_{\rm S}$) against a background radiation source (assumed to be the CMB)  of temperature $T_\gamma$ as a function of frequency $\nu$, given by 
\begin{equation}
\label{eq:T_b}
\begin{split}
\delta T_{b} (\nu)  & = \dfrac{T_{\rm S} - T_{\gamma}}{1 + z} (1 - e^{-\tau_{\nu_0}})\\
&\approx  27 x_{\textsc{\ion{H}{i}}}(1 + \delta_{\rm nl} ) \left( \dfrac{H}{dv_{r}/dr + H} \right) \left( 1 - \frac{T_{\gamma}}{T_{\rm S}} \right)\\
&\mathrm{\hspace{0.3cm}}\times \left( \frac{1+z}{10} \frac{0.15}{\Omega_{\rm M} h^{2}}\right) \left(\dfrac{\Omega_{\rm b} h^{2}}{0.023}\right)
\mathrm{ mK},
\end{split}
\end{equation}
where $\tau_{\nu_0}$ is the optical depth at the 21-cm transition frequency $\nu_0$, $x_{\hi}$ is the neutral hydrogen fraction, $\delta_{nl}$ is the fluctuations in the underlying dark matter density field, $H$ is the Hubble parameter, and $dv_{r}/dr$ is the line-of-sight component of the radial derivative of the peculiar velocity. All of the terms in equation \ref{eq:T_b} are evaluated at $z = \nu_{0}/\nu-1$.

\section{Fiducial simulation and mock observations}
\label{sec:sense}
In this section, we describe our fiducial model as well as the mock observations which are used to forecast constraints on our astrophysical model parameters.
\subsection{Fiducial model parameters}
The $\genpsim$ simulation from \cite{Balu2022} was calibrated against  existing observables including UV LFs from \cite{Bouwens2015} and \cite{Bouwens2021}, and stellar mass functions from  \cite{Song2016} and \cite{Stefanon2021}. The reionisation history was calibrated using the Thomson scattering optical depth of free electrons to CMB photons from the \cite{Planck2018}  \citep[see Fig. 3 \& 4 of][]{Balu2022}. 
\begin{figure*}
		\includegraphics[width=\textwidth]{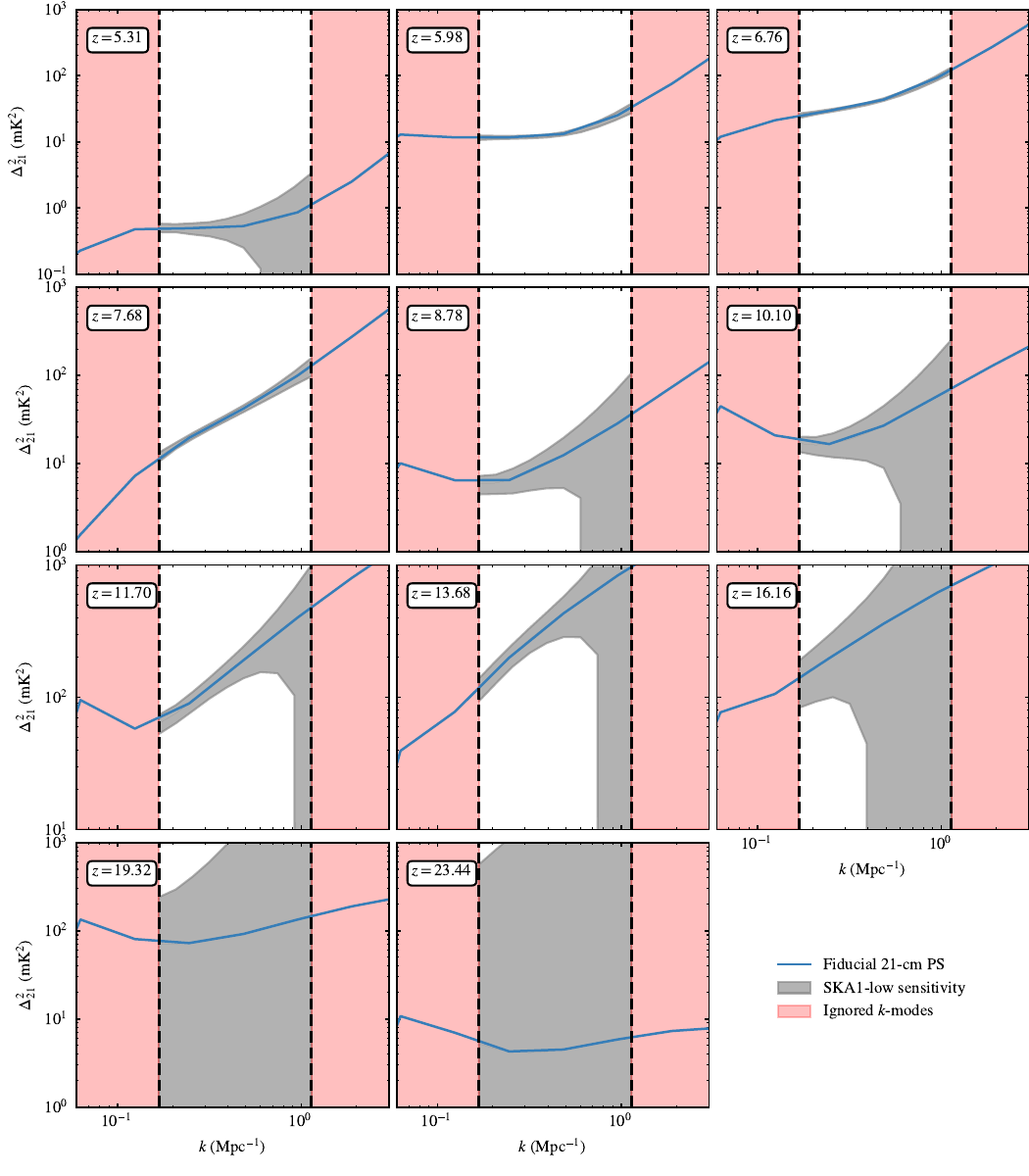}
        \caption{
        The blue curve shows the dimensionless 21-cm PS $\Delta_{21}^{2}$ (mK$^{2}$) from our fiducial model. 
        The grey-shaded region represents the sensitivity (including both thermal and cosmic variance noise) to the 21-cm PS for a 1000 h observation with the upcoming SKA1-low. We use the `moderate' foreground removal case from \citealt{Pober2014} effectively ignoring (pink-shaded region) all the $k$-modes falling within the 21-cm foreground wedge (setting a $k_{\rm min}=0.16$ Mpc$^{-1}$). The $k_{\rm max}=1.4$ Mpc$^{-1}$ is set by a combination of the spatial scales resolved by SKA1-low (set by the longest baseline in our model for the core) as well as scales we trust not to be dominated by the Poisson noise of the sources.
        }
        \label{fig:PS_fid_noise}
\end{figure*}
\begin{figure*}
		\centering
		\includegraphics[width=\textwidth]{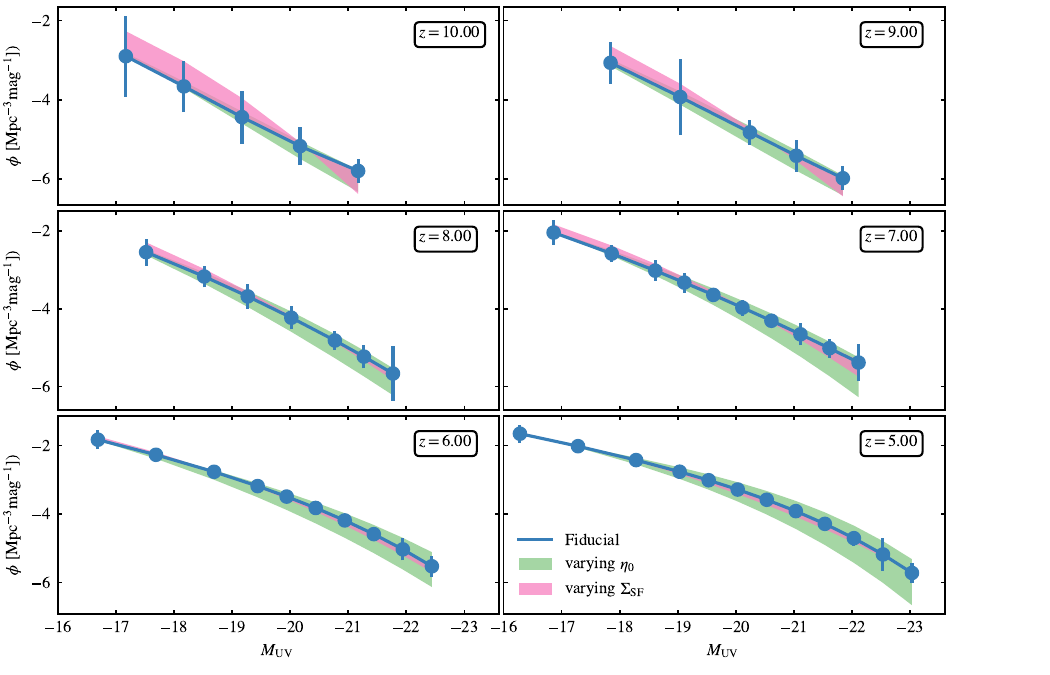}\vspace{-4mm}
        \caption{
        \label{fig:LFs} 
        The solid blue curve shows the fiducial UV LFs from our calibrated simulation. The uncertainties on our UV LFs are equivalent to the fractional uncertainties from \citealt{Bouwens2021}. The shaded regions show the variation in the UV LFs as we vary the critical mass normalisation ($\Sigma_{\rm SF}$; pink) and the efficiency of gas reheating due to SNe feedback ($\eta_0$; green) respectively. These parameters are varied by an order of magnitude about their fiducial values to explore their impact on the UV LFs and clearly demonstrate that including UV LFs in our Fisher matrix will improve our forecasted constraints as the range exceeds the 1-$\sigma$ observational uncertainty.
        }
\end{figure*}

The fiducial values of our 8 astrophysical model parameters are given in the fourth column of Table \ref{table:params}. These parameters can be divided into four groups of two in the order they are shown in the table, having direct control on the X-ray properties (\lxraygal \& $E_0$), the ionising UV photon escape fraction (\fnorm \& $\alpha_{\rm esc}$), star formation ($\Sigma_{\rm SF}$ \& $\alpha_{\rm SF}$), and the supernova feedback ($\varepsilon_0$ \& $\eta_0$) of the galaxies. 

\subsection{21-cm power spectra}
The frequency dependence of the 21-cm signal imparts a line-of-sight evolution to the 21-signal. To account for this, \meraxes produces a 21-cm light-cone by stitching together (linearly) interpolated $\delta T_b$ co-evolution grids. Following \cite{Brad_lightcone}, we subdivide our 21-cm light-cone into cubical grids of a size equivalent to our simulation volume ($1024^3$) and calculate the spherically averaged dimensionless 21-cm PS in each (shown as the solid blue curve in Fig. \ref{fig:PS_fid_noise}). We obtain 11 redshift chunks spanning $z\in[5,24]$, and assign redshift values to them corresponding to the mid-point of each chunk. 

\subsection{Modelling observational noise}

The sensitivity of a radio interferometer to the 21-cm PS can be divided into two components: (1) thermal noise, which is important on the small scales and (2) sample (cosmic) variance, prominent on the large scales. The total noise power $\sigma [ \Delta_{21}^2(k)]$ is given by adding these two in quadrature:
\begin{equation}\label{eq:error_full}
\left[\frac{1}{\sigma [\Delta_{21}^2(k)]}\right]^2=\sum_i\left(\frac{1}{\Delta_{\mathrm{N}, \mathrm{i}}^2+\Delta_{21}^2}\right)^2,
\end{equation}
where $\Delta_{\rm N}^2(k)$ is the thermal noise given by \citep{Morales2005, McQuinn2006, Parsons2012}:
\begin{equation}
\Delta_{\mathrm{N}}^2(k) \approx X^2 Y \frac{k^3}{2 \pi^2} \frac{\Omega^{\prime}}{2 t} T_{\mathrm{sys}}^2,
\end{equation}
where $X$ and $Y$ relate the bandwidths and solid angles to the comoving distance to the source, $\Omega^{'}$ is a factor dependant on the beam of the telescope \citep[see][]{Parsons2014}, $t$ is the integration time for the mode $k$, and $T_{\rm sys}$ is the system temperature given by $T_{\rm sys} = T_{\rm sky} + T_{\rm rec}$ with $T_{\rm sky}$ and $T_{\rm rec}$ being the sky and receiver temperature respectively. The quadrature addition and the form of the cosmic variance noise in equation \ref{eq:error_full} assume that the errors are Gaussian distributed which is reasonable for the relevant scales in this work \citep[see][]{Yuxiang_tale2, YuxiangLya2021, Prelogovic2023}.

In this work, we focus on a future observation of the 21-cm PS by the SKA. We limit our attention to the upcoming first phase of SKA -- the so-called SKA1-low\footnote{See the official \href{https://www.skatelescope.org/wp-content/uploads/2012/07/SKA-TEL-SKO-DD-001-1_BaselineDesign1.pdf}{SKA1 System Baseline Design document} for further details.}. We only include the stations in the `Central Area' of the SKA1-low, resulting in 296 stations  of diameter 35 m distributed across a circular area with 1.7 km diameter, we calculate the interferometer's sensitivity to the 21-cm PS using the \sense\footnote{\href{https://github.com/steven-murray/21cmSense}{www.github.com/steven-murray/21cmSense}}  \textsc{python} package \citep{Pober2013, Pober2014}.  We assume an observational campaign corresponding to 6 hours per night for 180 days, i.e.  a total of $1080$ hours and sky temperature $T_{\rm sky}$ to be dominated by galactic synchrotron emission \citep{T_Sky} scaling with frequency $\nu$ as $T_{\rm sky} = 60 (\nu/300 \rm{MHz})^{-2.55} \rm{K}$. Additionally, assuming that the telescope reflects $10$ per cent of the response to the sky \citep{Pober2014}, we set $T_{\rm rec} = 40 \rm{mK} + 0.1 T_{\rm sky}$. We combine partially coherent baselines to improve power spectrum sensitivity and use the `moderate' foreground removal scenario of \sense wherein we avoid the modes that are contained within the so-called foreground wedge \citep{Datta2010} extending $k_{\parallel}=0.1 h$ Mpc$^{-1}$ beyond the horizon limit. As we ignore the modes that fall within the foreground wedge, we are limited to $k_{\rm min} = 0.16$ Mpc$^{-1}$ for our analysis. We also have an upper limit of $k_{\rm max} = 1.4$ Mpc$^{-1}$ which arises from a combination of  both the spatial scales that are probed by the SKA1-low (set by the longest baseline we consider in our modelling of the array), as well as Poisson, shot noise from our simulation. Figure \ref{fig:PS_fid_noise} shows the PS noise (grey shaded region) that is generated for our fiducial 21-cm PS. The red-shaded region shows $k$-modes that are ignored in this work. 

\subsection{Luminosity Functions}

In addition to exploring the possible constraints on our fiducial model from the 21-cm PS we also consider the improvements that are achievable with a joint analysis of both the 21-cm PS and the UV LFs. 

In addition to updates on the SNe feedback (see section \ref{sec:SNe}), \cite{Dragons19} explored different implementations of dust attenuation in \meraxes. Here, we adopt the parameterisation for the UV optical depth that depends on the dust-to-gas ratio \citep{Charlot2000} of the galaxies. The model differentiates between a short-lived birth cloud of the stars as well as the interstellar medium (ISM) of the galaxy. Photons are absorbed by both the birth cloud (albeit for a short while until it gets depleted by star formation) and the ISM. The free parameters of \meraxes -- particularly the ones impacting the galaxy physics -- have been calibrated \citep[see][]{Dragons19, Balu2022} to their fiducial values against infrared excess (IRX)-$\beta$, UV LFs, and stellar mass functions at $z>5$.

The blue curves in Fig. \ref{fig:LFs} are the UV LFs from our calibrated simulation. The error bars are determined by multiplying our simulated UV LF data by the corresponding fractional uncertainty on each data point from the \cite{Bouwens2021} observational data. In addition to the 21-cm PS, we thus have 6 UV LFs from $z\in[5, 10]$.

\section{Forecasts using Fisher Matrices}\label{sec:FM}
To place quantitative constraints on the model parameters we use the Fisher information matrix \citep[$\mathbf{F}_{i j}$;][]{Tegmark1997, Albrecht2009}. For any set of observations, the Fisher matrix provides the best possible constraints on the parameters of an assumed model. An implicit assumption is that the errors on these parameters are Gaussian and that the observational data points are statistically independent. In this limit, by the Cramer-Rao theorem, the covariance matrix ($\mathbf{C}_{i j}$) of the parameters is given by the inverse of the Fisher matrix
\begin{equation}
    \mathbf{C}_{i j} = \mathbf{F}^{-1}_{i j}.
\end{equation}
The $1\sigma$ error on the $i$th parameter is given by the corresponding term on the diagonal of the covariance matrix i.e. $\mathbf{C}_{i i}$. Another relevant property of the Fisher matrices is their additive nature enabling one to do joint analyses of a different set of observations. This is achieved by adding the corresponding Fisher matrices together before calculating the joint $\mathbf{C}_{i j}$. 

Given the likelihood function $\mathcal{L}$  (the probability of the data given the model parameters  $\theta$) $\mathbf{F}_{i j}$\ is given by
\begin{equation}
\mathbf{F}_{i j} \equiv-\left\langle\frac{\partial^2 \ln \mathcal{L}}{\partial \theta_i \partial \theta_j}\right\rangle.
\end{equation}
For the present work, we compute $\mathbf{F}_{i j}$ as
\begin{equation}
\label{eq:FM}
    \mathbf{F}_{i j} = \sum_{k, z} \frac{1}{\sigma^2(k, z)} \frac{\partial \Delta^2(k, z)}{\partial \theta_i} \frac{\partial \Delta^2(k, z)}{\partial \theta_j},
\end{equation}
where $\Delta^2(k,z)$ is the dimensionless 21-cm PS, $\sigma(k,z)$ is the observational uncertainty on a measured $\Delta^2(k,z)$ (see equation \ref{eq:error_full}), and the summation is over all the $k$-modes and $z$-bins. Fisher matrices are thus sensitive to the derivatives of the 21-cm PS, with a larger value indicating increased sensitivity of the statistic to the model parameter. This translates into tighter constraining power on the model parameter. Parameters with a similar structure for the derivatives, as a function of $k$, will be degenerate \citep{Pober2014}.  In Fig. \ref{fig:ps_deriv} of appendix \ref{sec:appendix}, we show the noise $[\sigma(k,z)]$ weighted derivatives of the 21-cm PS with respect to each parameter we study in this work. Figure \ref{fig:PS_fid_noise} make it evident that the noise levels corresponding to a $\sim1000$ h observation with SKA1-low is lowest for the redshifts $z\lesssim8$. Thus, from equation \ref{eq:FM}, the constraining power will be dominated by these redshifts. Since the 21-cm PS is sensitive to different astrophysical parameters at different epochs \citep{BibleReview}, this will be reflected in our forecasted constraints. 

Except for $E_0$, $f_{\rm esc,0}$ and $\alpha_{\rm esc}$, all other parameters are varied in log-space as they can vary by at least an order of magnitude. For calculating the Fisher matrix  we vary each parameter by $\pm4$ per cent around its fiducial value\footnote{We explored different choices of the step-size and (visually) ensured the convergence of the derivatives.}. In order to further avoid numerical artefacts while computing the derivatives, we fitted the mock 21-cm PS with a 5$^{\rm th}$-order polynomial in log-space and interpolated the values for k-bins evenly spaced in log$_{10}(k)$\footnote{Polynomial fitting with different orders as well as in linear scales were explored to ensure convergence.}.

The $\mathbf{F}_{i j}$ corresponding to the UV LFs are computed by appropriately modifying equation \ref{eq:FM}.  Figure \ref{fig:lf_deriv} shows the derivatives of the UV LFs with respect to our model parameters after being weighted by the error $[\sigma(M_{\rm UV}, z)]$. We note here that while constructing the Fisher matrix corresponding to the UV LFs, we do not consider the ones varying the X-ray properties (i.e. $E_0$ \& $\rm{log}_{10}(L_{\rm X}/SFR)$) as they do not have any impact on the galaxy properties which affect its luminosity\footnote{The primary impact of the X-ray photons is to set the spin temperature of the IGM gas. For scenarios with extremely high values of the X-ray luminosity, they can cause $\sim 10-15$ per cent impact on the ionisations of the \hi \citep{MesingerXRay}. In these extreme cases, the X-ray can provide some constraints from UV LFs via the baryon modifier $f_{\rm mod}$. We have visually checked the constraints and have found them to be negligible for our fiducial model.}.

The primary advantage of a Fisher matrix analysis is the associated computational efficiency as they are quick and easy to calculate. This is particularly important when a single model evaluation is too computationally expensive for an MCMC, as is the case for \meraxes\footnote{A single model evaluation takes $\sim16$ h on two 48-core nodes.}.

\section{Results}\label{sec:results}
\begin{figure*}
    \includegraphics[width=\textwidth]{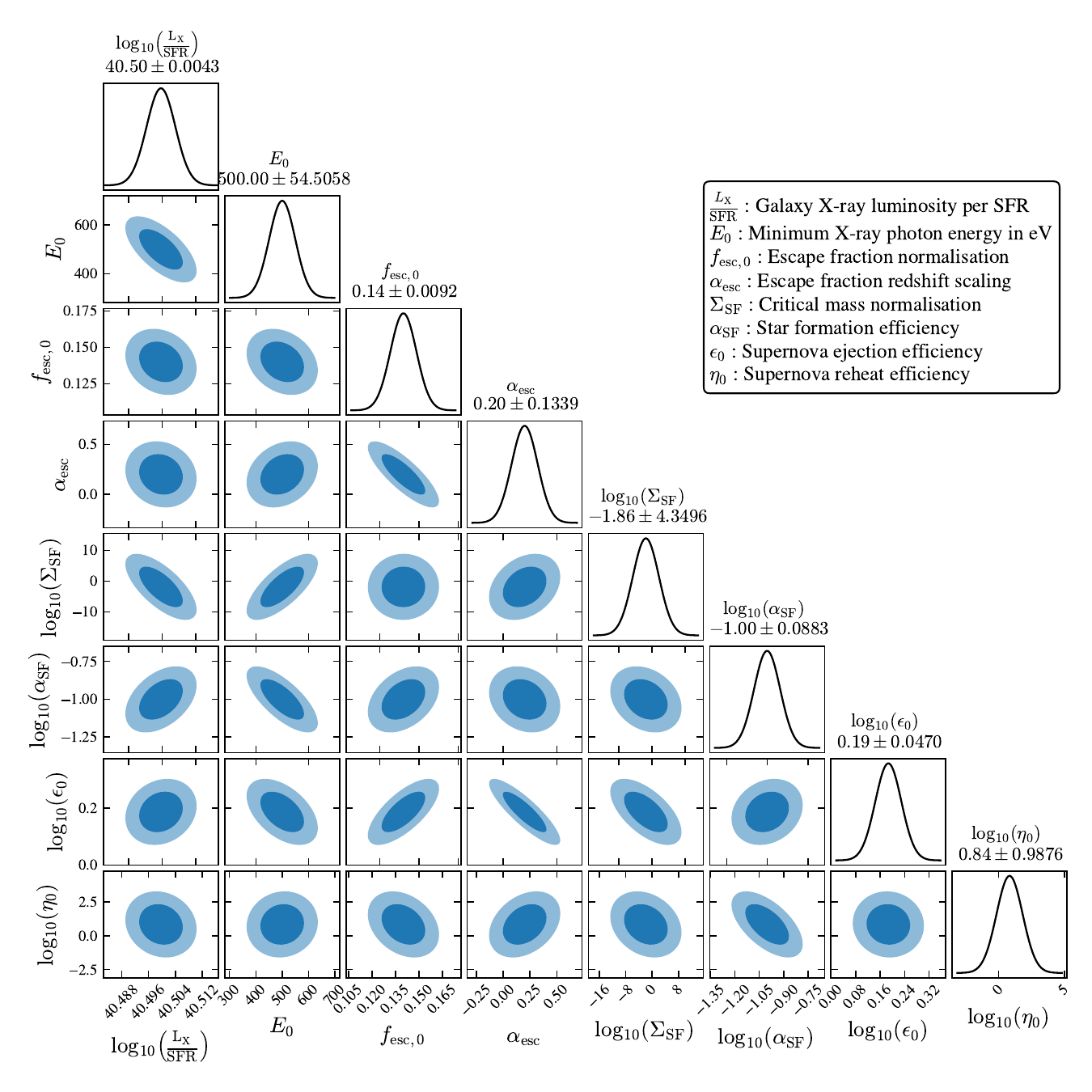}\vspace{-4mm}\caption{\label{fig:PS_only} 
    Constraints on our astrophysical model parameters from our Fisher matrix using a 1000 h mock observation with the SKA1-low. Dark (light) contours represent the 2-D marginalised confidence intervals and the diagonal subplots show the 1-D marginalised probability distribution functions of our parameters. }
\end{figure*}
\begin{figure*}
		\includegraphics[width=\textwidth]{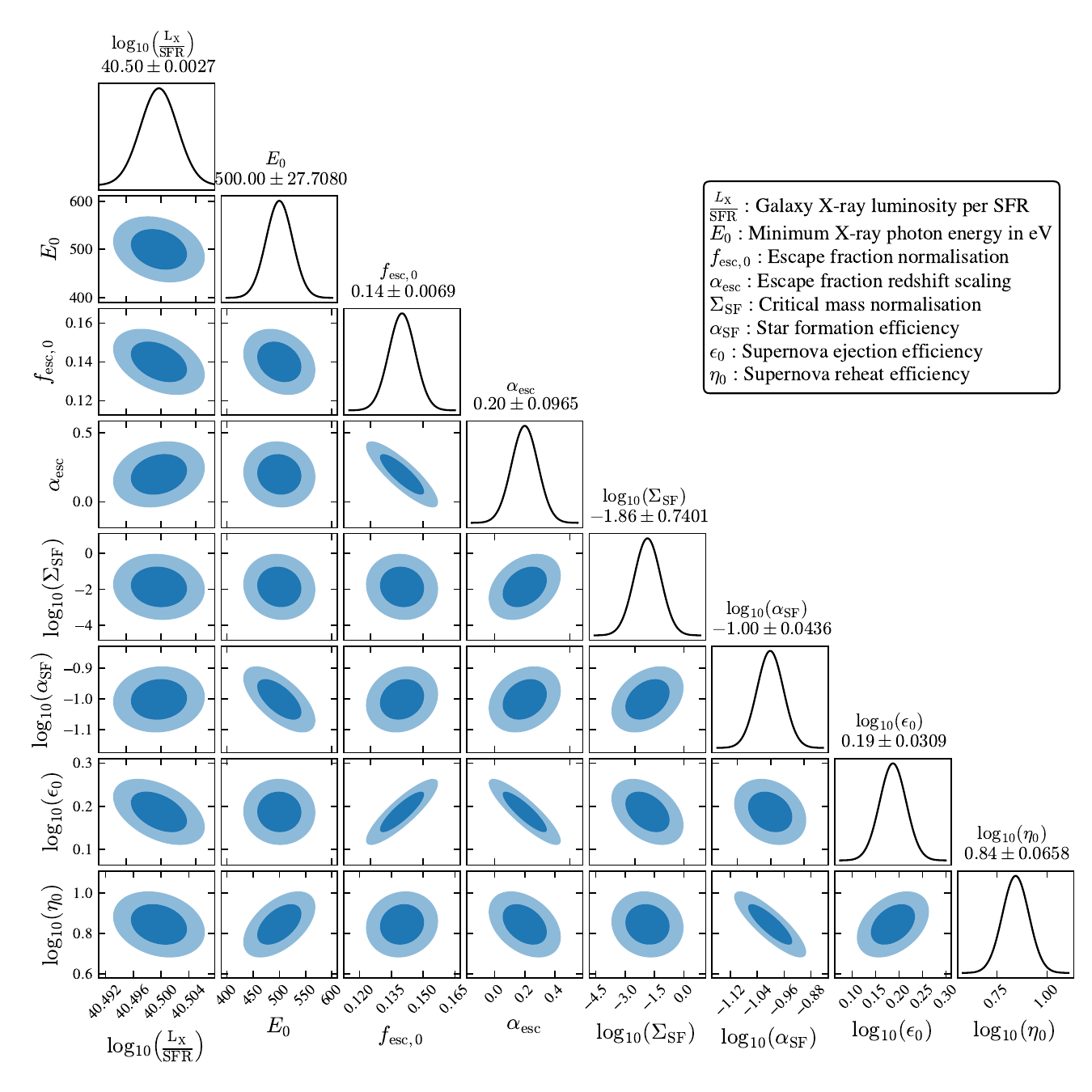}\vspace{-4mm}
		\caption{\label{fig:LFs_PS} Same as Fig. \ref{fig:PS_only} except that we have also added in constraints from the UV LFs. This removes some of the degeneracies between our parameters and results in tighter constraints.}
\end{figure*}
\begin{figure}
\includegraphics[width=\columnwidth]{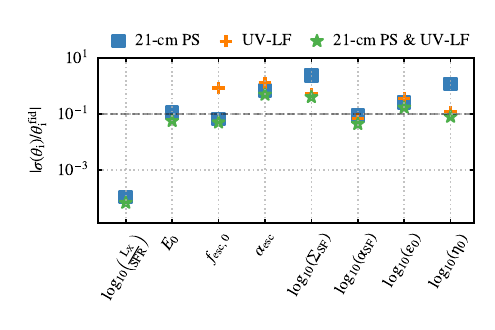}\vspace{-4mm}
\caption{\label{fig:fractional_error} Fractional uncertainties on our eight free parameters. The vertical axis is the $1-\sigma$ uncertainties divided by their fiducial values. The blue squares are from just the 21-cm PS, the orange crosses from just the UV LFs, and the green stars are from the joint analysis. While constructing the UV LFs Fisher matrix we do not consider the parameters controlling the X-ray luminosity of the galaxies, \lxraygal \& $E_0$. The grey dotted line shows the $10$ per cent fractional uncertainty limit .}
\end{figure}
In this section, we report the quantitative constraints available with a mock observation using just the 21-cm PS and also combining the 21-cm PS with the UV LFs.
\subsection{Forecasts from the 21-cm PS alone}
In Fig. \ref{fig:PS_only} we show the forecasts on our astrophysical parameter set from the 21-cm PS corresponding to a $\sim1000$ h observation with SKA1-low. The dark (light) contours correspond to the 1 (2)-$\sigma$  2-D marginalised confidence intervals for each astrophysical parameter and the diagonal subplots show the 1-D normalised marginal probability distribution function. The fifth column of table \ref{table:params} lists the 1-$\sigma$ uncertainty on our parameters. The fractional uncertainties, $|\sigma(\theta_i)/\theta^{fid}_i|$, are given within brackets and are shown as blue squares in Fig. \ref{fig:fractional_error}.

We obtain tight constraints ($\lesssim10$ per cent) for the X-ray parameters (luminosity \lxraygal \& minimum X-ray photon energy escaping galaxies $E_0$), escape fraction normalisation (\fnorm) and star formation efficiency ($\alpha_{\rm SF}$) while the SNe ejection efficiency ($\epsilon_0$) is constrained to $\sim25$ per cent. On the other hand, the critical mass normalisation ($\Sigma_{\rm SF}$) and the efficiency of SNe reheating ($\eta_0$) remain relatively unconstrained with $\sim234$ per cent and $\sim117$ per cent fractional $1\sigma$ uncertainties respectively. The relatively poor constraints on $\Sigma_{\rm SF}$ and $\eta_0$ are primarily because these parameters have negligible impact on the 21-cm PS. 

As previously mentioned, our parameter set forms four groups of two with each group controlling a different aspect of galaxy properties. 

\begin{enumerate} 
    \item \textit{X-ray parameters}: The 21-cm signal is very sensitive to the amount and energy of the X-ray photons in the early Universe. The 21-cm PS, therefore, provides tight constraints on the X-ray parameters, $E_0$ ($\sim11$ per cent) and \lxraygal ($\sim 10^{-2}$ per cent), in our model.  These forecasts are consistent with similar works in the literature \citep[e.g.][]{CMMC2, Park2019, 21cmfish}, which is unsurprising given they have the same X-ray implementation despite the differences in the source modelling (i.e. SAM).
    
    \item \textit{UV escape fraction}: $f_{\rm esc}$ is more sensitive to the normalisation \fnorm as opposed to the  redshift power-law exponent ($\alpha_{\rm esc}$). This is also reflected in the derivative of the 21-cm PS with respect to these parameters (green and pink curves in Fig. \ref{fig:ps_deriv}). We thus  forecast tighter constraints on \fnorm (6.55 per cent) compared to $\alpha_{\rm esc}$ (66.93 per cent).
    
    \item \textit{Star formation}: Among the star formation parameters, the efficiency of star formation ($\alpha_{\rm SF}$) is highly constrained relative to the critical mass normalisation ($\Sigma_{\rm SF}$). The primary impact of $\Sigma_{\rm SF}$ is to establish a critical mass threshold needed for the stars to start forming. This parameter thus has only a secondary role in the stellar mass content of galaxies as once a galaxy has accreted enough mass, the amount of stellar material is controlled  by $\alpha_{\rm SF}$ (see section \ref{sec:SF}). Equation \ref{eq:SF_law}  also shows that these two parameters enter our model as a multiplicative term of the form $(1-\Sigma_{\rm SF})\alpha_{\rm SF}$ adding to their degeneracy.
    
    \item \textit{Supernovae feedback}: Compared to the star formation parameters, we find that the SNe parameters are less constrained, with $\sim25$ per cent for the SNe ejection efficiency $\epsilon_0$ while the efficiency of reheating by the SNe $\eta_0$ is only constrained to  $\sim117$ per cent. Similar to $\Sigma_{\rm SF}$, the low constraints on these parameters are due to the fiducial model values being such that the 21-cm signal is relatively insensitive to these processes.
\end{enumerate}

\subsection{Joint forecasts using the 21-cm PS and the UV LFs}

We next add in the Fisher matrix corresponding to the UV LFs from 6 redshifts $\in[5,10]$. As the UV LFs functions are sensitive to parameters that do not have a pronounced impact on the ionisation morphology, the addition of the UV LFs into the analysis helps improve the overall constraints on the astrophysical model \citep{Park2019}. We show the impact of varying the two least constrained parameters when using just the 21-cm PS, $\Sigma_{\rm SF}$ and $\eta_0$, on the UV LFs in Fig. \ref{fig:LFs}. The green and pink shaded regions are from varying $\eta_0$ and $\Sigma_{\rm SF}$ respectively, by an order of magnitude about their fiducial values (we point out that these two parameters are positive by definition). The variation in the UV LFs is larger than the 1-$\sigma$ errors emphasising that including the UV LFs in our Fisher analysis will provide additional constraining power. As previously mentioned, we do not vary the X-ray parameters, \lxraygal \& $E_0$, while computing the UV LF Fisher matrix. 

Figure \ref{fig:LFs_PS} shows the constraints from the joint analysis of the 21-cm PS and UV LFs, and the last column of Table \ref{table:params} lists the forecasted 1-$\sigma$ uncertainty along with the fractional uncertainties for our model parameters. The green stars in Fig. \ref{fig:fractional_error} are the fractional uncertainties for the joint analysis of both the 21-cm PS and the UV LFs, and the orange crosses represent the same from just the UV LFs\footnote{In Fig. \ref{fig:LFs_only} we show the constraints from the UV LFs alone.}. Figures \ref{fig:PS_only} \& \ref{fig:fractional_error} demonstrate that combining the UV LFs into the analyses helps to improve the constraints. This is most notable for parameters having a direct impact on the stellar mass of the galaxies i.e. the ones related to star formation and supernova feedback as UV LFs are more sensitive to these parameters compared to the 21-cm PS. This results in significant improvements in $\Sigma_{\rm SF}$ (from $\sim234$ to $\sim40$ per cent) and $\eta_0$ (from $\sim117$ to $\sim8$ per cent) for the joint case of both the 21-cm PS and the UV LFs as the degeneracies between stellar properties and escape fraction can be weakened. This then translates into improvements in other parameters.

The relatively larger improvement in $\Sigma_{\rm SF}$ compared to $\alpha_{\rm SF}$ following the inclusion of information from the UV LFs stems from the breaking of parameter degeneracies. In Fig. \ref{fig:PS_only} (21-cm PS only), $\Sigma_{\rm SF}$ is poorly constrained (highlighted by the large uncertainties), and $\Sigma_{\rm SF}$ and $\alpha_{\rm SF}$ are anti-correlated. In Fig. \ref{fig:LFs_only} (UV LFs only), $\Sigma_{\rm SF}$ is now more tightly constrained and positively correlated with $\alpha_{\rm SF}$, indicating the majority of $\Sigma_{\rm SF}$'s constraining power comes from the UV LF. Thus, the inclusion of UV LFs provide considerable improvements to the constraining power of $\Sigma_{\rm SF}$. For  $\alpha_{\rm SF}$ on the other hand, it is almost equally well constrained both by the 21-cm PS (Fig. \ref{fig:PS_only}) and the UV LFs (Fig \ref{fig:LFs_only}). Thus combining the two observables provides modest improvements (dominated by the change in correlation).

\subsection{Discussion of parameter degeneracies}
Even though Fisher matrices implicitly make simplifying assumptions, such as Gaussian likelihoods and Gaussian errors, they nevertheless give useful insights. In this section, we explore the correlations and degeneracies among the model parameters. For brevity, we focus only on the results from the joint analysis of the UV LFs and the 21-cm PS (Fig \ref{fig:LFs_PS}), noting that the trends are similar for the 21-cm PS alone (Fig. \ref{fig:PS_only}). 

\begin{enumerate}
    \item \textit{X-ray parameters}: \lxraygal anti-correlates with the UV escape fraction normalisation (\fnorm) and the SNe ejection efficiency ($\epsilon_0$). $E_0$ shows an anti-correlation with the star formation efficiency ($\alpha_{\rm SF}$) and positively correlates with the SNe reheat efficiency ($\eta_0$). 
    
    These correlations can be understood from the impact of the X-rays on the EoR. X-rays can contribute, albeit a small fraction relative to the UV photons, to the ionisation of the local IGM of a galaxy leading to photoionisation regulation of the amount of neutral gas available for accretion onto the galaxy and hence subsequent star formation. Thus increasing \lxraygal can be compensated by a decrease in the SNe feedback and/or the ionising UV escape fraction. $E_0$ sets the lowest energy for escaping X-rays from the galaxies. An increase in $E_0$, therefore, implies less IGM ionisation by the X-rays.
    
    \item \textit{UV escape fraction}:  The correlation among $\alpha_{\rm esc}$ and \fnorm is expected given the definition of the UV escape fraction ($f_{\rm esc}$; see equation \ref{eq:f_esc}). The UV escape fraction, $f_{\rm esc}$, directly influences the local IGM ionisation state, and  hence the amount of baryonic infall into the galaxy for star formation. The amount of cold gas available for star formation is influenced by the SNe ejection efficiency which  sets the amount of gas removed from the galaxy. An increase in $f_{\rm esc}$ therefore should be accompanied by a decrease in the SNe feedback.
    
    In light of this, the relatively strong positive correlation of \fnorm  with the supernova ejection efficiency $\epsilon_0$ is interesting. This can be understood from the behaviour of the UV escape fraction (i.e. the combination of \fnorm and $\alpha_{\rm esc}$) and the other physical processes.  The positive correlation amongst the SNe parameters ($\eta_0$ \& $\epsilon_0$), as can be seen from equations (\ref{eq:m_new} - \ref{eq:E_hot}) and Fig. \ref{fig:LFs_only}, also plays a contributing factor. $\alpha_{\rm esc}$ is also anti-correlated with both $\eta_0$ and $\epsilon_0$. It is therefore a combination of the very weak correlation of \fnorm with $\eta_0$ and the anti-correlation with $\alpha_{\rm esc}$ that results in its positive correlation with $\epsilon_0$.

    The comparatively tighter correlation of the escape fraction parameters with the SNe parameters compared to the X-ray parameters reflects the relative importance of the UV photons over the X-ray photons in the IGM ionisation state. 
    
    \item \textit{Star formation}: As noted, we report correlations between the star formation efficiency $\alpha_{\rm SF}$ and the minimum X-ray photon energy $E_0$. 
    
    The strong correlation between $\alpha_{\rm SF}$ and SNe reheat efficiency $\eta_0$ is expected as they are two of the parameters impacting the stellar mass in a galaxy.

    On the other hand, the correlation between $\alpha_{\rm SF}$ and  the critical mass normalisation $\Sigma_{\rm SF}$ is surprisingly weak given equation (\ref{eq:SF_law}). This is because of the relatively small value of $\Sigma_{\rm SF}$. $\Sigma_{\rm SF}$ fixes the critical mass of the cold gas needed for star formation to commence in a galaxy. The small value of the parameter is justified on the physical grounds that a larger value will result in lower available gas for star formation.
    
    \item \textit{Supernovae feedback}: For completeness, we again point out the correlations between the SNe ejection ($\epsilon_0$) and reheat ($\eta_0$) efficiencies, the strong correlations of $\epsilon_0$ with \fnorm \& $E_0$, and that of $\eta_0$ with $\alpha_{\rm SF}$.
    
\end{enumerate}

\section{conclusion}\label{sec:conclusion}
Using the semi-analytic galaxy formation model \meraxes we perform a Fisher matrix analysis to forecast the constraints on physical properties of galaxy formation and reionisation that will be available from future 21-cm PS observations. Specifically, we use the 210 $\oneh$ Mpc cosmological simulation resolving all atomically cooled haloes (\genpsim of \citealt{Balu2022}) down from $z=20$. 

SAMs are unique in enabling computationally efficient exploration of the underlying parameter space corresponding to the complex astrophysics of galaxy formation and evolution. We focused on 8 free parameters in our model that directly impact the X-ray luminosity, UV escape fraction, star formation rate and SNe feedback of the galaxies. We constructed a mock observation of the 21-cm PS, focusing on a 1000 hr observation with the forthcoming SKA1-low. The observational uncertainty was modelled assuming foreground wedge avoidance using the \textsc{python} package \sense. Using the Fisher matrix formalism, we find that 4 (5) out of the 8 parameters can be constrained to within $\lesssim10$ ($\lesssim 50$) per cent using the 21-cm PS alone from the EoR. Specifically, we forecast constraints on our X-ray parameters (the luminosity \lxraygal and the minimum energy of the photon escaping the early galaxies $E_0$) that are comparable to similar works in the literature. We also forecast tight constraints on parameters controlling the star formation efficiency ($\alpha_{\rm SF}$) as well as the normalisation of the UV escape fraction (\fnorm) of the early galaxies. On the other hand, SNe feedback parameters remain largely unconstrained reflecting that the 21-cm PS is relatively insensitive to them. 
The complex astrophysics of the early galaxy formation and evolution is captured in the degeneracies and correlations among the model parameters. Of particular interest are the correlations among the UV escape fraction and SNe feedback, and those between the star formation efficiency and X-ray parameters of the galaxies.

To improve the overall constraining power of our analysis we added the Fisher matrix corresponding to the UV LFs from redshifts $z\in[5, 10]$. This results in an improvement in all of our parameter forecasts, most notably the critical mass normalisation $\Sigma_{\rm SF}$ and the SNe reheat efficiency $\eta_0$. This is not surprising as these parameters primarily control the stellar mass content of the galaxies to which the UV LFs are very sensitive. Incorporating the UV LFs into the analyses results in 5 of our parameters being constrained to $\lesssim10$ per cent and all 8 of them being to within $\lesssim50$ per cent. Our forecasts illustrate that detailed observations of reionisation with the SKA will be valuable in constraining the astrophysics of the early galaxies.

\section*{Acknowledgements}

We thank the referee for their detailed comments which improved the quality of this manuscript. SB thanks Yuxiang Qin for the helpful discussions which assisted this work. This research was supported by the Australian Research Council Centre of Excellence for All Sky Astrophysics in 3 Dimensions (ASTRO 3D), through project \#CE170100013. Part of this work was performed on the OzSTAR national facility at the Swinburne University of Technology. The OzSTAR program partially receives funding from the Astronomy National Collaborative Research Infrastructure Strategy (NCRIS) allocation provided by the Australian Government. This research also made use of resources from the National Computational Infrastructure (NCI Australia), another NCRIS-enabled capability supported by the Australian Government.

\textit{Software citations}:
This research relies heavily on the \textsc{python} \citep{PYTHON} open source community, in particular, \textsc{numpy} \citep{NUMPY}, \textsc{matplotlib} \cite{MATPLOTLIB}, \textsc{scipy} \citep{SCIPY}, \textsc{h5py}, \textsc{jupyter} \citep{Jupyter}, and \textsc{pandas} \citep{Pandas}.
\section*{Data Availability}
The data underlying this article will be shared on reasonable request to the corresponding author.
\bibliographystyle{mnras}
\bibliography{references}
\bsp	
\label{lastpage}
\appendix
\section{Derivatives of the 21-cm PS}
\label{sec:appendix}
\begin{figure*}
	\begin{minipage}{\textwidth}
		\centering
		\includegraphics[width=\textwidth]{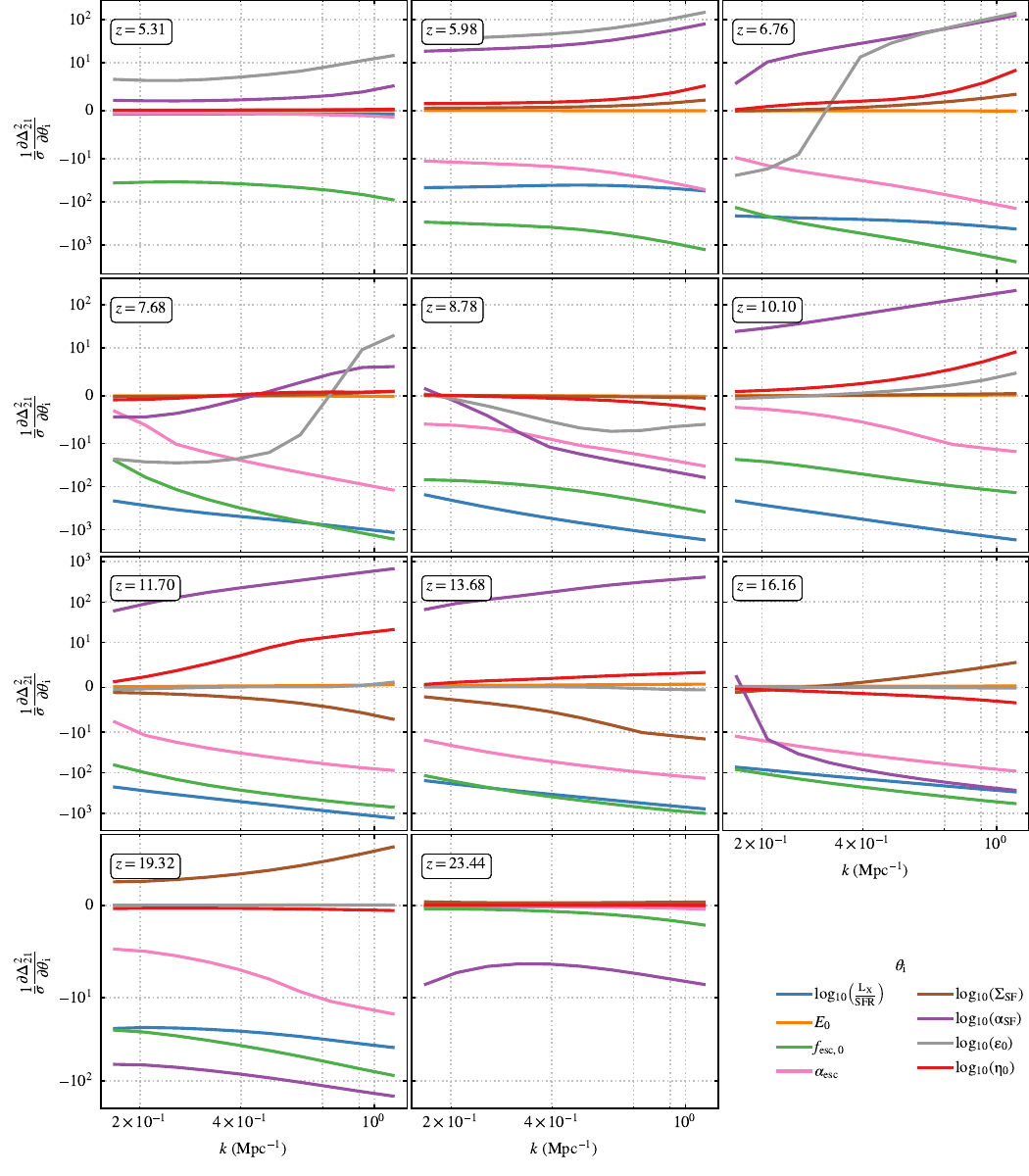}\vspace{-4mm}
		\caption{\label{fig:ps_deriv} The derivatives of the 21-cm PS [$\partial\Delta_{21}^2/\partial\theta_i$] with respect to the parameters varied in this work. The derivatives are weighted by the noise levels [$\sigma(k,z)$]. The vertical axes is linear $\in[-10, 10]$ and log-scale otherwise.}
	\end{minipage}
\end{figure*}

In this section, we show the derivatives of our primary statistics i.e. the 21-cm PS (Fig. \ref{fig:ps_deriv}). These derivatives are computed by perturbing the $\genpsim$ simulation about the fiducial model, one parameter $\theta_i$ at a time. These are then used to compute the Fisher matrix (see equation \ref{eq:FM}). Parameters corresponding to similarly shaped derivatives will be degenerate. Figure \ref{fig:ps_deriv} shows these derivatives [$\partial\Delta_{21}^2/\partial\theta_i$] weighted by the noise powers [$\varepsilon(k,z)$], i.e. $\frac{1}{\varepsilon}\frac{\partial\Delta_{21}^2}{\partial\theta_i}$. We have employed a 'symmetric log' scaling on the vertical axis in the figure with the scale being linear $\in[-10, 10]$ and log-scale otherwise.

\section{Adding in the UV LFs}

\begin{figure*}
	\begin{minipage}{\textwidth}
		\centering
		\includegraphics[width=\textwidth]{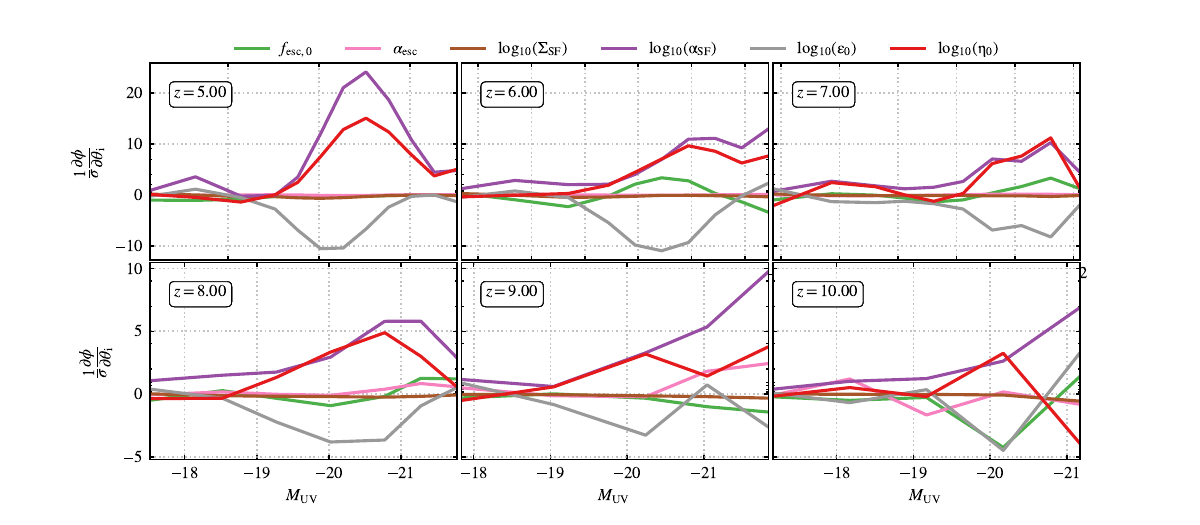}\vspace{-4mm}
		\caption{\label{fig:lf_deriv} The derivatives of the UV LFs [$\partial\phi/\partial\theta_i$] with respect to the parameters varied in this work. The derivatives are weighted by the noise levels [$\sigma(M_{\rm UV},z)$].}
	\end{minipage}
\end{figure*}

This section shows the analysis using only the UV LFs. Figure \ref{fig:lf_deriv} shows the derivatives of the UV LFs, computed by perturbing the $\genpsim$ simulation about the fiducial model, one parameter $\theta_i$ at a time. These are then used to compute the Fisher matrix. We do not vary the X-ray parameters, \lxraygal and $E_0$, for the UV LFs analysis as they do not have an impact on the UV LFs. Parameters corresponding to similarly shaped derivatives will be degenerate. Figure \ref{fig:lf_deriv} shows these derivatives [$\partial\phi/\partial\theta_i$] weighted by the noise powers [$\sigma(M_{\rm UV},z)$], i.e. $\frac{1}{\sigma}\frac{\partial\phi}{\partial\theta_i}$.

\begin{figure*}
    \includegraphics[width=\textwidth]{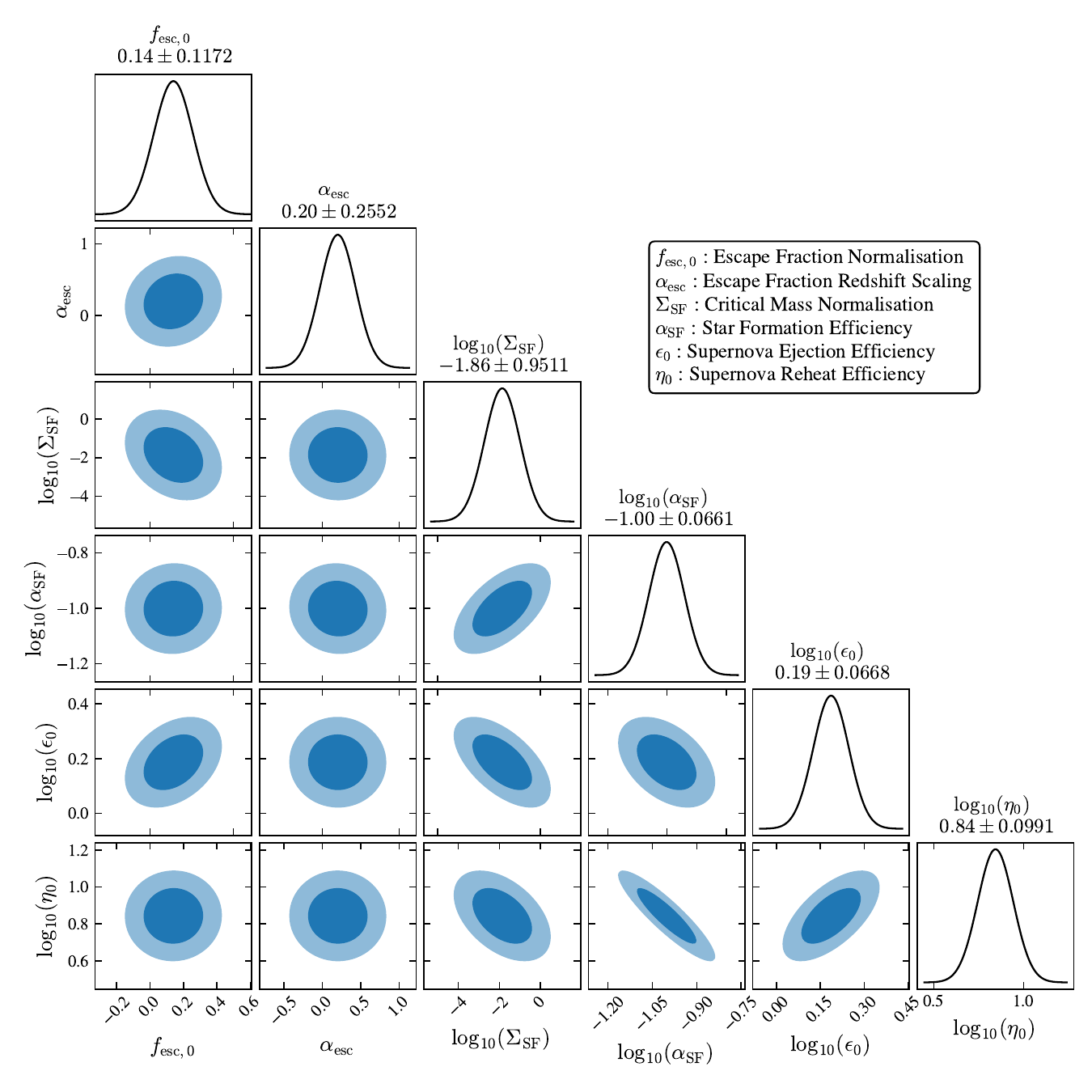}\vspace{-4mm}\caption{\label{fig:LFs_only} 
    Constraints on our astrophysical model parameters from our Fisher matrix analysis using only the UV LFs. Dark (light) contours represent the 2-D marginalised confidence intervals and the diagonal subplots show the 1-D marginalised probability distribution functions of our parameters. Note that while computing the UV LFs Fisher matrices we do not vary the X-ray parameters (see text for more details).}
\end{figure*}

In Figure \ref{fig:LFs_only} we show the forecasted constraints from the UV LFs Fisher matrix. The dark (light) contours correspond to the 1 (2)-$\sigma$  2-D marginalised confidence intervals for each astrophysical parameter and the diagonal subplots show the 1-D normalised marginal probability distribution function.

\label{sec:appendix2}

\end{document}

%% file: new_commands.tex
\newcommand{\lya}{\ifmmode\mathrm{Ly}\alpha\else{}Ly$\alpha$\fi\:}
\newcommand{\lyb}{\ifmmode\mathrm{Ly}\beta\else{}Ly$\beta$\fi}
\newcommand{\igm}{\ifmmode\mathrm{IGM}\else{}IGM\fi}
\newcommand{\lae}{\ifmmode\mathrm{LAE}\else{}LAE\fi}
\newcommand{\h}{\ifmmode\mathrm{H}\else{}H\fi}
\newcommand{\hi}{\ifmmode\mathrm{H\,{\scriptscriptstyle I}}\else{}H\,{\scriptsize I}\:\fi}
\newcommand{\heii}{\ifmmode\mathrm{He\,{\scriptscriptstyle II}}\else{}He\,{\scriptsize II}\fi}
\newcommand{\heiii}{\ifmmode\mathrm{He\,{\scriptscriptstyle III}}\else{}He\,{\scriptsize III}\fi}
\newcommand{\ciii}{\ifmmode\mathrm{C\,{\scriptscriptstyle III]}}\else{}C\,{\scriptsize III]}\fi}
\newcommand{\oiii}{\ifmmode\mathrm{O\,{\scriptscriptstyle III}}\else{}O\,{\scriptsize III}\fi}
\newcommand{\aliii}{\ifmmode\mathrm{Al\,{\scriptscriptstyle III}}\else{}Al\,{\scriptsize III}\fi}
\newcommand{\mgii}{\ifmmode\mathrm{Mg\,{\scriptscriptstyle II}}\else{}Mg\,{\scriptsize II}\fi}
\newcommand{\fe}{\ifmmode\mathrm{Fe}\else{}Fe\fi}
\newcommand{\nv}{\ifmmode\mathrm{N\,{\scriptscriptstyle V}}\else{}N\,{\scriptsize V}\fi}
\newcommand{\niv}{\ifmmode\mathrm{N\,{\scriptscriptstyle IV]}}\else{}N\,{\scriptsize IV]}\fi}
\newcommand{\cii}{\ifmmode\mathrm{C\,{\scriptscriptstyle II}}\else{}C\,{\scriptsize II}\fi}
\newcommand{\civ}{\ifmmode\mathrm{C\,{\scriptscriptstyle IV}}\else{}C\,{\scriptsize IV}\fi}
\newcommand{\siv}{\ifmmode\mathrm{Si\,{\scriptscriptstyle IV}}\else{}Si\,{\scriptsize IV}\fi}
\newcommand{\siii}{\ifmmode\mathrm{Si\,{\scriptscriptstyle II}}\else{}Si\,{\scriptsize II}\fi}
\newcommand{\siiii}{\ifmmode\mathrm{Si\,{\scriptscriptstyle III]}}\else{}Si\,{\scriptsize III]}\fi}
\newcommand{\ovi}{\ifmmode\mathrm{O\,{\scriptscriptstyle VI}}\else{}O\,{\scriptsize VI}\fi}
\newcommand{\sioiv}{\ifmmode\mathrm{Si\,{\scriptscriptstyle IV}\,\plus O\,{\scriptscriptstyle IV]}}\else{}Si\,{\scriptsize IV}\,+O\,{\scriptsize IV]}\fi}
\newcommand{\Msun}{\rm M_\odot}

\newcommand{\sense}{$\textsc{21cmsense}$\:}

\newcommand{\appropto}{\mathrel{\vcentre{
			\offinterlineskip\halign{\hfil$##$\cr
				\propto\cr\noalign{\kern2pt}\sim\cr\noalign{\kern-2pt}}}}}

\newcommand{\lsim}{\mathrel{\rlap{\lower4pt\hbox{\hskip1pt$\sim$}}
        \raise1pt\hbox{$<$}}}
\newcommand{\gsim}{\mathrel{\rlap{\lower4pt\hbox{\hskip1pt$\sim$}}
        \raise1pt\hbox{$>$}}}
\newcommand{\Rom}[1]{\uppercase\expandafter{\romannumeral #1}}
\newcommand{\rom}[1]{\lowercase\expandafter{\romannumeral #1}}

\newcommand{\hii}{\ion{H}{ii}\:}

\newcommand{\oneh}{h^{-1}}
\newcommand{\dforest}{\textsc{DarkForest}\:}

\newcommand{\genpsim}{\textsc{L210\_AUG}\:}

\newcommand{\cmfast}{\textsc{21cmFAST}\:}

\newcommand{\meraxes}{\textsc{Meraxes}\:}

\newcommand{\lxraygal}{{$L_{\rm X}/\rm{SFR}$}\:}
\newcommand{\fnorm}{{$f_{\rm esc, 0}$}\:}